\documentclass[prb,showpacs,twocolumn,aps]{revtex4}
\usepackage{graphicx}
\newcommand{\be}{\begin{equation}}
\newcommand{\ee}{\end{equation}}
\newcommand{\bea}{\begin{eqnarray}}
\newcommand{\eea}{\end{eqnarray}}
\newcommand{\up}{\uparrow}
\newcommand{\dn}{\downarrow}
\newcommand{\bwt}{\begin{widetext}}
\newcommand{\ewt}{\end{widetext}}
\begin{document}
\title{Dynamics of a magnetic moment induced by a 
spin-polarized current
}
\author{Wonkee Kim and F. Marsiglio}
\affiliation{
Department of Physics, University of Alberta, Edmonton, Alberta,
Canada, T6G~2J1}
\begin{abstract}
Effects of an incoming spin-polarized current on a magnetic moment
are explored.
We found that the spin torque occurs only when
the incoming spin changes as a function of time inside of the magnetic
film. 
This implies that some modifications
are necessary in a phenomenological model where the coefficient of the
spin torque term is a constant, and the coefficient is determined by dynamics
instead of geometrical details.
The precession of the magnetization reversal 
depends on the incoming energy of electrons in the spin-polarized current.
If the incoming energy is smaller than the interaction energy,
the magnetization does not precess while reversing its direction. 
We also found that the relaxation time associated with the
reversal depends on the incoming energy. 
The coupling between an incoming spin and a magnetic moment
can be estimated by measuring the relaxation time.
\end{abstract}
\maketitle

\section{introduction}

Tremendous attention has been paid to the dynamics of magnetization in
recent years because this problem is of fundamental importance in understanding
magnetism and because the problem is of interest to technological applications
in magnetic devices.\cite{hillebrands}
One of intriguing features of magnetization motion is spin transfer
from a spin-polarized current to a magnetization of a ferromagnetic film,
theoretically proposed by Slonczewski\cite{slonczewski} 
and Berger\cite{berger}, and later experimentally
verified.\cite{myers,katine}
Since this spin transfer mechanism was first conceptualized, 
many studies\cite{bazaliy,sun,waintal,stile,zhang} have 
been performed on this phenomenon. However, the dynamics of a magnetic moment
driven by a spin-polarized current has not been fully explored. 

In this paper we investigate the current-driven precession and reversal 
of a magnetic moment. This is done  quantum mechanically
using a simple Hamiltonian, without introducing an external magnetic field.
In this way one can easily distinguish contributions from the current
from those induced by an externally applied field. To this end,
we describe the motion of a magnetic moment in the lab frame where
details of the magnetization reversal are best illustrated. 
Since dynamics of a magnetic moment can be formally described 
in the local moment frame and such a description may also give
some intuition about the dynamics, we examine an interaction between 
a spin-polarized current and a magnetic moment in the local frame at the
Hamiltonian level in section (II).
Then dynamics in the lab frame 
is illustrated in section (III). In this section one can see details
of the dynamics such as under what conditions
the motion of the magnetization can be non-precessional or the relaxation time
associated with the reversal
is a minimum. These phenomena have not been explored in the literature so far.
Section (IV) is devoted to discussions about the adiabatic approximation
used to describe the motion of a magnetic moment, and we close with a summary.

\section{Formalism in the local moment frame}

To describe effects of an incoming spin current on a magnetic moment
${\bf M}$, as in Ref.\cite{bazaliy}
one can choose a frame $(X'Y'Z')$, where ${\hat z}'$ is parallel
to ${\bf M}$. Such a frame is called the local magnetic moment frame.
Extensive work on ferromagnetism in the local moment frame has been
done in Ref.\cite{koreman}
An advantage of this frame is that it is trivial to diagonalize
an interaction between an incoming spin ${\bf s}$ and a magnetic moment:
$-2J_{H}{\bf M}\cdot{\bf s}$, where $J_{H}$ is the coupling.

Let us start with a simple Hamiltonian relevant to the interaction:
\bwt
\be
H=\int{}dx\left[
\psi^{+}_{\alpha}(x)\left(-\frac{\nabla^2}{2m}\right)\psi_{\alpha}(x)
-2J_{H}{\bf M}(x)\cdot{\bf s}
+V(x)\psi^{+}_{\alpha}(x)\psi_{\alpha}(x)\right]
\label{old_H}
\ee
\ewt
where $\psi^{+}_{\alpha}(x)$ creates an electron with a spin $\alpha$
at $x$, $m$ is the electron mass and
$V(x)$ is an impurity potential. 
The magnitude of the magnetic moment is $M_{0}$, which remains unchanged.
The electron spin can be represented
as $s^{i}=(1/2)\psi^{+}_{\alpha}\sigma^{i}\psi_{\beta}$, where $\sigma^{i}$
is a Pauli matrix with $i=x,y$ and $z$. 
We assume that the magnetic moment
is determined by a localized electron $\Psi$ so that the kinetic part
of the localized electron is not included in the Hamiltonian.

Suppose a local magnetic moment ${\bf M}({\bf x})$ points in the direction
$(\theta,\;\phi)$ at ${\bf x}$ as seen in Fig.~1. Then,  
a local rotation ( or coordinate transformation to the local
moment frame) is introduced: 
$
\psi_{\alpha}(x)=U_{\alpha\beta}(x)\chi_{\beta}(x)\;,
$
where
\bwt
\be
U(x)=\left(\begin{array}{cc}
  \cos(\theta/2)e^{-i\phi/2}&-\sin(\theta/2)e^{-i\phi/2}\\
  \sin(\theta/2)e^{i\phi/2}&\cos(\theta/2)e^{i\phi/2}
\end{array}\right)\;.
\ee
In terms of $\chi(x)$, the Hamiltonian can be written as
\be
H=\int{}dx\left[
\frac{1}{2m}\nabla\left(\chi^{+}_{\beta}U^{+}_{\beta\alpha}\right)
\cdot\nabla\left(U_{\alpha\gamma}\chi_{\gamma}\right)-
J_{H}\chi^{+}_{\beta}U^{+}_{\beta\alpha}({\bf M}\cdot\sigma_{\alpha\mu})
U_{\mu\nu}\chi_{\nu}+V(x)\chi^{+}_{\alpha}\chi_{\alpha}\right]
\ee
Since the interaction term in the Hamiltonian is
diagonalized in this $\chi(x)$ basis, we obtain
\be
H=H_{0}+\int{}dx\left[
{\bf A}_{\alpha\beta}\cdot{\bf j}_{\alpha\beta}
+
A^{(0)}_{\alpha\beta}\rho_{\alpha\beta}\right]\;,
\label{new_H}
\ee
where
\bea
H_{0}&=&\int{}dx\left[
\chi^{+}_{\alpha}(x)\left(-\frac{\nabla^2}{2m}\right)\chi_{\alpha}
-J_{H}M_{0}\chi_{\alpha}\sigma^{z}_{\alpha\beta}\chi_{\beta}
+V(x)\chi^{+}_{\alpha}(x)\chi_{\alpha}(x)\right]
\nonumber\\
{\bf j}_{\alpha\beta}&=&\frac{1}{2im}\left[\chi^{+}_{\alpha}\nabla\chi_{\beta}
-(\nabla \chi^{+}_{\alpha})\chi_{\beta}\right]\;,
\hskip 0.5cm 
{\bf A}_{\alpha\beta}=-iU^{+}_{\alpha\gamma}(\nabla U_{\gamma\beta})\;,
\nonumber\\
A^{(0)}_{\alpha\beta}&=&\frac{1}{2m}(\nabla U^{+}_{\alpha\gamma})\cdot
(\nabla U_{\gamma\beta})\;,
\hskip 0.5cm \mbox{and} \hskip 0.5cm
\rho_{\alpha\beta}=\chi^{+}_{\alpha}\chi_{\beta}\;.
\nonumber
\eea
\ewt
After diagonalizing the interaction, we have an extra term
$H'=\int{}dx\left[
{\bf A}_{\alpha\beta}\cdot{\bf j}_{\alpha\beta}
+
A^{(0)}_{\alpha\beta}\rho_{\alpha\beta}\right]$ 
in Eq.~(\ref{new_H})
instead of off-diagonal terms
of the interaction in Eq.~(\ref{old_H}). Using the
explicit form of $U(x)$, we can calculate
vector potentials $A^{(0)}_{\alpha\beta}$ and ${\bf A}_{\alpha\beta}$.
This was the route followed in Ref.\cite{bazaliy}, which led to a monopole-like
term in the energy. Those authors attributed the spin torque term
to this new vector potential, which is purely geometrical.

Here we follow a different route, since we are interested in a
simpler case, where the magnetization is {\it not } a function of position.
Thus, in our case of a single-domain ferromagnet, the extra term shown 
above will disappear because $\nabla U=0$. Instead, our spin torque will
be present due to the dynamics of the coupled spin-moment system. In addition,
we will not require an assumption regarding the magnitude of $J_{H}$ in
order to proceed, and
we
will utilize an impurity potential for convergence purposes which is
otherwise irrelevant to the spin transfer as in Ref.\cite{bazaliy}

\begin{figure}[tp]
\begin{center}
\includegraphics[height=2.6in,width=3.6in]{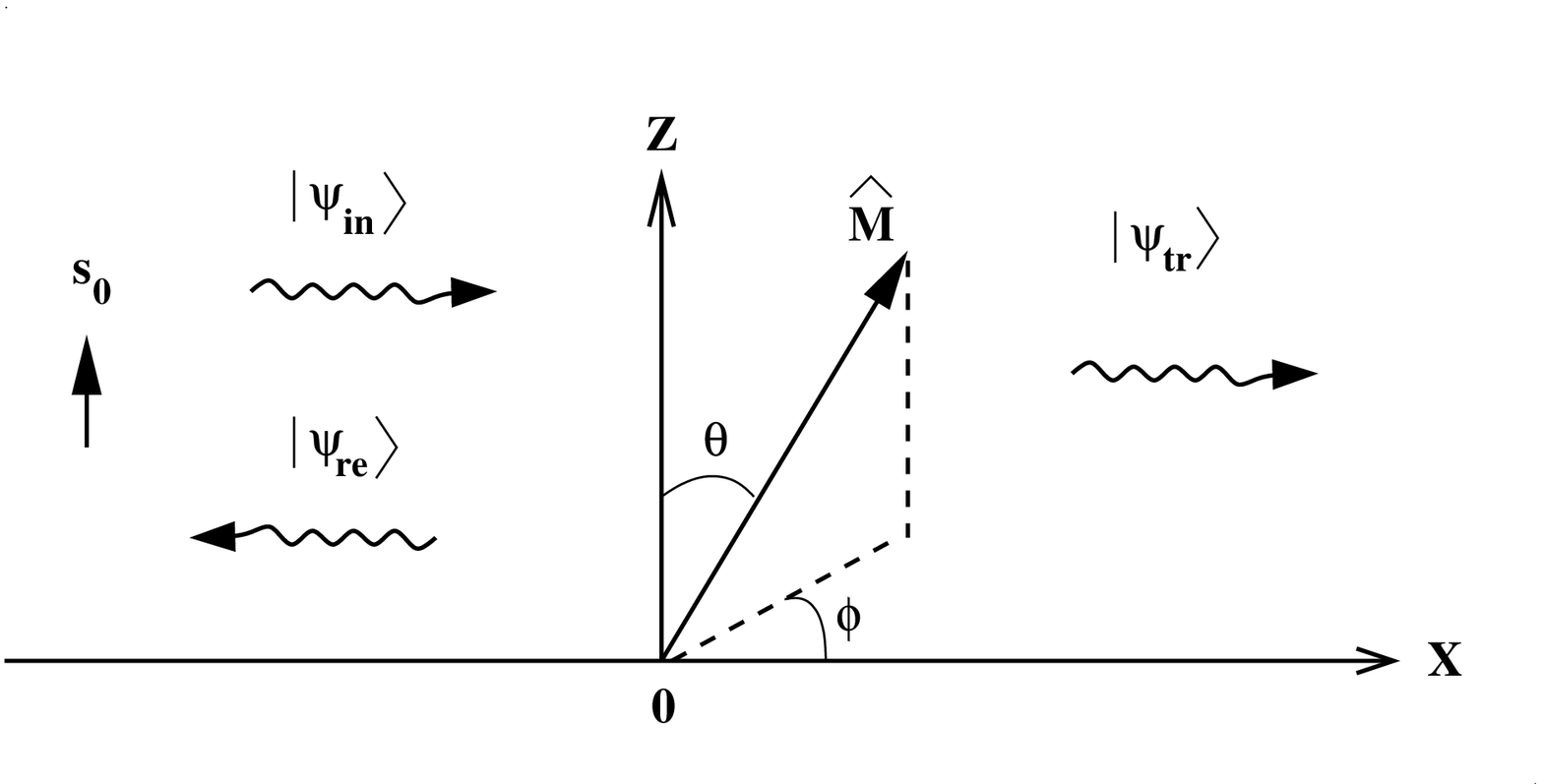}
\caption{Geometry of a quantum mechanical problem associated with
the spin transfer. The incoming electron to the positive $X$ axis
are spin-polarized along ${\hat z}$ axis.
The ferromagnet surface is at $x=0$ and parallel to $YZ$ plane.
The direction of the magnetic moment is defined by $\theta$ and $\phi$,
which are functions of time $t$. The ferromagnet is assumed
to be sufficiently thick.
}
\end{center}
\end{figure}

\section{Dynamics of a magnetic moment in the lab frame}

A disadvantage of the description in the local moment frame is
that the precession of the magnetic moment cannot be seen;
in other words, a precessional reversal of the magnetic
moment cannot be distinguished from a plain reversal. 
Since our goal
in this paper is to investigate the dynamics of the magnetic moment
as mentioned in the introduction, we describe the motion of
the magnetic moment in the lab frame. The geometry of our problem is shown
in Fig.~1. We assume a single-domain ferromagnet in the $YZ$ plane
for simplicity and
consider the Hamiltonian Eq.(\ref{old_H}). 
The incoming spin is along ${\hat z}$ and the direction of the
magnetic moment is defined by $\theta(t)$ and $\phi(t)$, which vary as
functions of time $t$.

The equation of motion for the magnetic moment ${\bf M}$ can be obtained
quantum mechanically: 
$d{\bf M}/dt=i\left[H,{\bf M}\right]$.
Since $M^{i}=(1/2)\gamma_{0}
\Psi^{+}_{\alpha}\tau^{i}_{\alpha\beta}\Psi_{\beta}$,
where $\Psi$ and $\tau^{i}$ are the operator and
a Pauli matrix for localized electrons, respectively, and $\gamma_{0}$ 
is the gyromagnetic ratio,
the equation becomes
\be
\frac{d{\bf M}}{dt}=2\gamma_{0}J_{H}\left({\bf M}\times{\bf s}\right)\;.
\label{mdot}
\ee
To analyze this equation we consider ${\bf M}$ as a classical vector
and take ${\bf s}$ as its expectation value over the ferromagnet.
If we decompose ${\bf s}$ into a parallel ${\bf s}_{\parallel}$ and
a perpendicular ${\bf s}_{\perp}$ component to ${\bf M}$, we
know that only ${\bf s}_{\perp}$ contributes to the equation.
We can express ${\bf s}_{\perp}$ using any unit vector. Let us choose,
for the unit vector,
the initial direction of the incoming spin ${\hat s}_{0}={\hat z}$. Then
\be
{\bf s}_{\perp}=s_{\perp}({\hat M}\times{\hat s}_{0})+
s'_{\perp}\left[{\hat M}\times({\hat s}_{0}\times{\hat M})\right]\;,
\label{s_perp}
\ee
where
$s_{\perp}=\frac{{\hat s}_{0}\cdot({\bf s}\times{\hat M})}
{1-({\hat s}_{0}\cdot{\hat M})^{2}}$
and
$s'_{\perp}=\frac{{\hat s}_{0}\cdot{\bf s}-({\hat s}_{0}\cdot{\hat M})
({\bf s}\cdot{\hat M})}{1-({\hat s}_{0}\cdot{\hat M})^{2}}$.
Using Eq.~(\ref{s_perp}), we can rewrite Eq.~(\ref{mdot}) as follows:
\be
\frac{d{\bf M}}{dt}=-2\gamma_{0}J_{H}s_{\perp}{\bf M}\times({\hat s}_{0}
\times{\hat M})+2\gamma_{0}J_{H}s'_{\perp}({\bf M}\times{\hat s}_{0})\;.
\ee
As we can see in the above equation, the first term on the right hand side
gives
the spin torque while the second term causes a precession of the magnetic 
moment. We emphasize that the spin torque occurs only when ${\bf s}(t)$ 
changes
as a function of time $t$. If ${\bf s}$ remains 
parallel to ${\hat s}_{0}$, then
$s_{\perp}$ vanishes and no spin torque takes place. In this instance,
the effect of a spin is the same as that of an external magnetic field
along ${\hat z}$ and the magnetic moment precesses. In a phenomenological
model,\cite{sun} the spin torque is represented by
${\bf M}\times({\hat s}_{0}\times{\hat M})$ with a proportional constant.
However, a time dependence of $s_{\perp}$ is crucial as we emphasized.
We also should stress that $s_{\perp}$ and $s'_{\perp}$ are 
determined by dynamics, not
geometrical details as in Ref.\cite{zhang}

To evaluate the expectation value of ${\bf s}$, we need to solve
the Schr{\" o}dinger equation for the Hamiltonian Eq.~(\ref{old_H}). 
Basically, the equation is one-dimensional because of translational symmetry
in the $YZ$ plane.
We choose the direction
of the polarized spin to be ${\hat z}$. 
Then, an incoming wave function $|\psi_{in}\rangle$
with a momentum ${\bf k}$ or an energy $\epsilon=k^{2}/2m$ is 
$|+\rangle e^{ikx}$, where $|+\rangle$ is the spin-up state in the lab frame.
We need to consider a normalization factor $C$ for $|\psi_{in}\rangle$.
Since this wave function describes an electron beam, $|C|^{2}$ is
the number of electrons $N_{e}$ per unit length in one dimension. Intuitively,
the more electrons are bombarded into the ferromagnet, 
the stronger is the effect of spin transfer. 
We thus expect the time scale for the reversal to scale inversely with
$N_e$ (the more the number of electrons, the faster the moment responds).
Similarly, the time scale will be proportional to the magnitude of the
local spin, $S_{local}\;(=M_{0}/\gamma_{0})$ (the
larger the moment, the longer it will take to reverse it). 

The reflected $(|\psi_{re}\rangle)$
and transmitted $(|\psi_{tr}\rangle)$ wave functions are eigenstates 
$|\chi_{\up}\rangle$ and $|\chi_{\dn}\rangle$ of the
interaction $2J_{H}{\bf M}\cdot{\bf s}=J_{H}{\bf M}\cdot{\bf\sigma}\;$; namely,
$J_{H}{\bf M}\cdot{\bf\sigma}|\chi_{\up}\rangle=J_{H}M_{0}|\chi_{\up}\rangle$
and 
$J_{H}{\bf M}\cdot{\bf\sigma}|\chi_{\dn}\rangle=-J_{H}M_{0}|\chi_{\dn}\rangle$.
Therefore,
\be
|\psi_{re}\rangle=\Bigl[
R_{\up}|\chi_{\up}\rangle\langle\chi_{\up}|+\rangle+
R_{\dn}|\chi_{\dn}\rangle\langle\chi_{\dn}|+\rangle\Bigr] e^{-ikx}
\label{psi_re}
\ee
while
\be
|\psi_{tr}\rangle=T_{\up}|\chi_{\up}\rangle\langle\chi_{\up}|+\rangle
e^{ik_{\up}x}+
T_{\dn}|\chi_{\dn}\rangle\langle\chi_{\dn}|+\rangle e^{ik_{\dn}x}\;,
\label{psi_tr}
\ee
where $k_{\up}=\sqrt{k+2mJ_{H}M_{0}}$ and $k_{\dn}=\sqrt{k-2mJ_{H}M_{0}}$
as depicted in Fig.~2. If the energy of the incoming electron is less than
$J_{H}M_{0}$, $k_{\dn}=i\kappa_{\dn}$ becomes pure imaginary where 
$\kappa_{\dn}=\sqrt{2mJ_{H}M_{0}-k}$,
and its corresponding
wave function decays exponentially; $e^{-\kappa_{\dn}x}$.
\begin{figure}[tp]
\begin{center}
\includegraphics[height=2.6in]{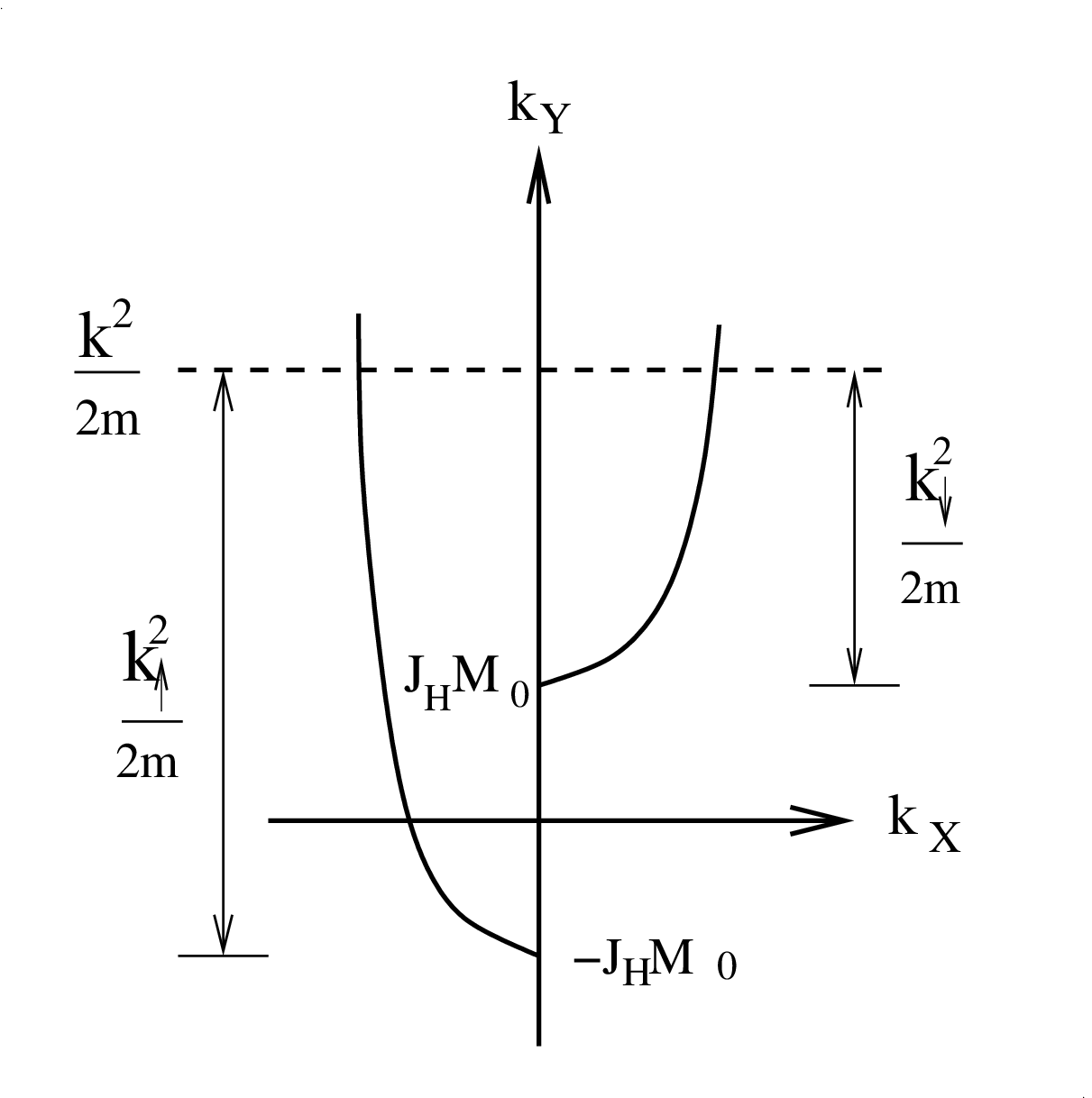}
\caption{An energy band and relations among $k^{2}/2m$,
$k^{2}_{\up(\dn)}/2m$, and $J_{H}M_{0}$. In this figure,
it is assumed that $k^{2}/2m > J_{H}M_{0}$.
}
\end{center}
\end{figure}

For $x<0$, $|\psi(x<0)\rangle=|\psi_{in}\rangle+|\psi_{re}\rangle$ and
for $x>0$, $|\psi(x>0)\rangle=|\psi_{tr}\rangle$. The coefficients
$R_{\up(\dn)}$ and $T_{\up(\dn)}$ are determined by matching conditions
of wave functions and their derivatives at $x=0$:
\be
R_{\up(\dn)}=\frac{k-k_{\up(\dn)}}{k+k_{\up(\dn)}}
\hskip 0.5cm\mbox{and}\hskip 0.5cm
T_{\up(\dn)}=\frac{2k}{k+k_{\up(\dn)}}\;.
\label{coeffs}
\ee
Note that we take $|\psi_{in}\rangle=|+\rangle e^{ikx}$ in the above
derivations. This means that
the number of electrons in the incoming beam $N_{e}$ is unity 
for simplicity;
however, when we numerically solve the equation of motion for a magnetic moment,
we can control this parameter.
In the Hamiltonian Eq.~(\ref{old_H}), we also have an impurity potential $V(x)$.
We shall introduce mean free paths $l_{\up}$ and $l_{\dn}$
for each channel due to the impurity, 
and as in Ref.\cite{berger}
they serve as convergence factors such as $e^{-x/l_{\up}}$
and $e^{-x/l_{\dn}}$ when we average
the expectation of ${\bf s}$ using $|\psi(x>0)\rangle$ over the ferromagnet.
We assume that the thickness of the ferromagnet $(L)$ is much larger than
the mean free paths: $L\gg l_{\up(\dn)}$.
One may wonder if the matching coefficients change when the convergence factors
are introduced. They do change as, for example, 
$k_{\up}\rightarrow k_{\up}+i/l_{\up}$; however, the conclusions we make later
remain unchanged as we verified.

Now we can calculate the expectation value of ${\bf s}$
within the ferromagnet; 
$\langle s^{i}\rangle=(1/2)\langle\psi_{tr}|\sigma^{i}|\psi_{tr}\rangle$
with $i=x\;,y$ and $z$. The average values of the expectation values
are evaluated as $\bar{\langle s^{i}\rangle}=(1/2)
\int^{L}_{0}dx\langle\psi_{tr}|\sigma^{i}|\psi_{tr}\rangle$.
After some straightforward algebra, we obtain for incoming energy
greater than $J_{H}M_{0}$
\bwt
\bea
\bar{\langle s^{x}\rangle}
&=&\frac{l_{\up}}{2}\mbox{Re}\left[\alpha^{*}\gamma\right]
+\frac{l_{\dn}}{2}\mbox{Re}\left[\beta^{*}\delta\right]
+\mbox{Re}\left[\frac{\beta^{*}\gamma+\delta^{*}\alpha}
{(1/l_{\up}+1/l_{\dn})-i(k_{\up}-k_{\dn})}\right]
\\
\bar{\langle s^{y}\rangle}
&=&-\frac{l_{\up}}{2}\mbox{Im}\left[\gamma^{*}\alpha\right]
-\frac{l_{\dn}}{2}\mbox{Im}\left[\delta^{*}\beta\right]
-\mbox{Im}\left[\frac{\delta^{*}\alpha-\beta^{*}\gamma}
{(1/l_{\up}+1/l_{\dn})-i(k_{\up}-k_{\dn})}\right]
\\
\bar{\langle s^{z}\rangle}
&=&\frac{l_{\up}}{4}\left(|\alpha|^{2}-|\gamma|^{2}\right)
+\frac{l_{\dn}}{4}\left(|\beta|^{2}-|\delta|^{2}\right)
+\mbox{Re}\left[\frac{\beta^{*}\alpha-\delta^{*}\gamma}
{(1/l_{\up}+1/l_{\dn})-i(k_{\up}-k_{\dn})}\right]\;,
\eea
\ewt
where
$\alpha=(1/2)T_{\up}(1+m_{z})$, $\beta=(1/2)T_{\dn}(1-m_{z})$,
$\gamma=(1/2)T_{\up}(m_{x}+im_{y})$, and 
$\delta=-(1/2)T_{\dn}(m_{x}+im_{y})$. Here ${\bf m}\;
(={\bf M}/\gamma_{0}S_{local})$ 
is the 
unit vector of the magnetic moment; namely,
$m_{z}=cos(\theta)$ and $m_{x}+im_{y}=\sin(\theta)e^{i\phi}$.

In our treatment, the incoming energy $\epsilon=k^{2}/2m$ is a control 
parameter and $J_{H}M_{0}$ is a scaling parameter. Experimentally,
$\epsilon$ can be controlled by adjusting the applied voltage while
$J_{H}M_{0}$ is uncontrollable because $J_{H}$ is a microscopic parameter.
If $\epsilon=\eta J_{H}M_{0}$, then $k^{2}_{\up}/2m=(\eta+1)J_{H}M_{0}$
and $k^{2}_{\dn}/2m=(\eta-1)J_{H}M_{0}$. Defining $k^{2}_{0}/2m=J_{H}M_{0}$,
$k_{\up}$ and $k_{\dn}$ can be written as $k_{\up}=\sqrt{\eta+1}\;k_{0}$
and $k_{\dn}=\sqrt{\eta-1}\;k_{0}$. Since
the current density is in energy units in 1D ($\hbar \equiv 1$),
using $j_{0}=k_{0}/m$ with one electron per unit length we can define
a dimensionless time $\tau=j_{0}t$, which will be 
used in the numerical calculations.
When $\eta < 1$, 
$k_{\dn}=i\sqrt{1-\eta}\;k_{0}$ as mentioned earlier. In this case
$\bar{\langle {\bf s}\rangle}$ changes to reflect $k_{\dn}=i\sqrt{1-\eta}\;k_{0}$.
We do not present equations for $\eta < 1$ here because the
derivation is parallel to
the above case and expressions are similar with those for $\eta > 1$.
Since we attribute the impurity potential to the mean free paths,
it is natural to assume $l_{\up}=l_{\dn}\equiv l$. We also introduce
a parameter $a=lk_{0}$. In the numerical calculations, we vary $a$ from $0.5$ to
$2$. Qualitative behaviors of ${\bf m}$ are not sensitive to the value of $a$.

A dimensionless equation of motion for the magnetic moment is
\be
\frac{d\bf m}{d\tau}=
\frac{N_{e}/2}{S_{local}}\left({\bf m}\times{\bf h}\right)\;,
\label{mdot2}
\ee
where
\bwt
\be
h_{i} = \frac{a}{4}|T_{\up}|^{2}(1+m_{z})m_{i}-
\frac{a}{4}|T_{\dn}|^{2}(1-m_{z})m_{i}+
\frac{4A_{i}/a-2(\sqrt{\eta+1}-\sqrt{\eta-1})B_{i}}
{{4}/{a^{2}}+\left(\sqrt{\eta+1}-\sqrt{\eta-1}\right)^{2}}
\ee
\ewt
with ($i=x,\;y$, and $z$)
\bea
A_{x}&=&
-\frac{1}{2}\left(\mbox{Re}\left[T_{\up}T^{*}_{\dn}\right]m_{x}m_{z}
+\mbox{Im}\left[T_{\up}T^{*}_{\dn}\right]m_{y}\right)
\nonumber\\
B_{x}&=&\frac{1}{2}\left(\mbox{Re}\left[T_{\up}T^{*}_{\dn}\right]m_{y}
-\mbox{Im}\left[T_{\up}T^{*}_{\dn}\right]m_{x}m_{z}\right)
\nonumber\\
A_{y}&=&
-\frac{1}{2}\left(\mbox{Re}\left[T_{\up}T^{*}_{\dn}\right]m_{y}m_{z}
-\mbox{Im}\left[T_{\up}T^{*}_{\dn}\right]m_{x}\right)
\nonumber\\
B_{y}&=&\frac{1}{2}\left(\mbox{Re}\left[T_{\up}T^{*}_{\dn}\right]m_{x}
+\mbox{Im}\left[T_{\up}T^{*}_{\dn}\right]m_{y}m_{z}\right)
\nonumber\\
A_{z}&=&
\frac{1}{2}\mbox{Re}\left[T_{\up}T^{*}_{\dn}\right]
\left(m^{2}_{x}+m^{2}_{y}\right)
\nonumber\\
B_{z}&=&\frac{1}{2}
\mbox{Im}\left[T_{\up}T^{*}_{\dn}\right]
\left(m^{2}_{x}+m^{2}_{y}\right)\;.
\nonumber
\eea
Clearly the factor $N_{e}/2S_{local}$ could be absorbed into the time
(already dimensionless).  Since
its effect is obvious, we set $N_{e}/2S_{local}=4$ for all our results.

We choose various values of $\eta$ between $0.25$ and $4$, and
show $m_{i}(\tau)$ vs. $\tau$ and a locus of 
${\bf m}$ in the $(m_{x},m_{y},m_{x})$
coordinate. For an initial condition of ${\bf m}$ we choose
$\theta_{0}=\pi/1.01$ and $\phi_{0}=\pi/4$ to see the magnetic moment reversal.
Because of a rotational symmetry, the initial value of $\phi$ is not important.
It is obvious that if $\theta_{0}=0$ or $\pi$, the spin polarized current has
no effect on ${\bf m}$. In Fig.~3(a), we 
show the locus (dotted curve) of ${\bf m}$ for $\eta=2$ and $a=1$, and 
plot $m_{i}(\tau)$ vs. $\tau$ in Fig.~3(b).
Thin circles define a uni-sphere.
Oscillations
in $m_{x}$ and $m_{y}$ imply precession of ${\bf m}$. For $\eta=2$, ${\bf m}$
shows a precessional reversal. On the other hand, for $\eta=0.5$ it has a plain reversal
without precession as we can see in Fig.~4(a) and (b). In this instance,
$m_{x}$ and $m_{y}$ do not show oscillations.
The precessional reversal takes place only when $\eta\ge1$. This remains true 
for $a=0.5$ or $2$. We plot these results in Fig.~5(a) and 5(b) for 
$\eta=0.25$ and $4$.
\begin{figure}[tp]
\begin{center}
\includegraphics[height=2.6in]{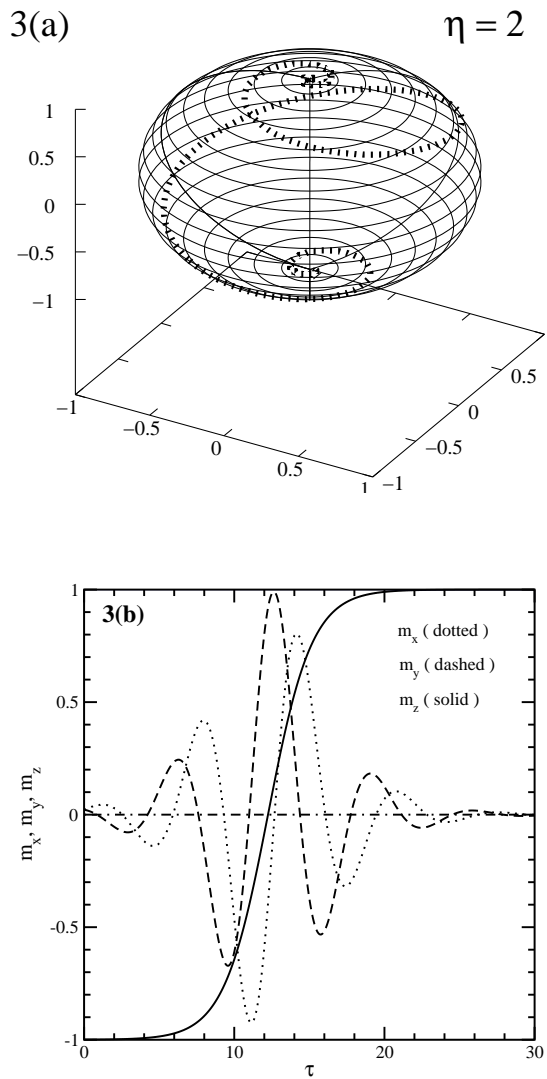}
\vskip 1.1cm
\includegraphics[height=2.6in]{fig3b.eps}
\caption{Precessional reversal of the magnetic moment for $\eta=2$ and $a=1$.
Fig.~3(a) shows the locus (dotted curve)
of ${\bf m}$ and Fig.~3(b) is for $m_{i}(\tau)$ vs. $\tau$.
The initial direction of ${\bf m}$ is given by $\theta_{0}=\pi/1.01$ and $\phi_{0}=\pi/4$.
Thin circles define a uni-sphere.
}
\end{center}
\end{figure}
\begin{figure}[tp]
\begin{center}
\includegraphics[height=2.6in]{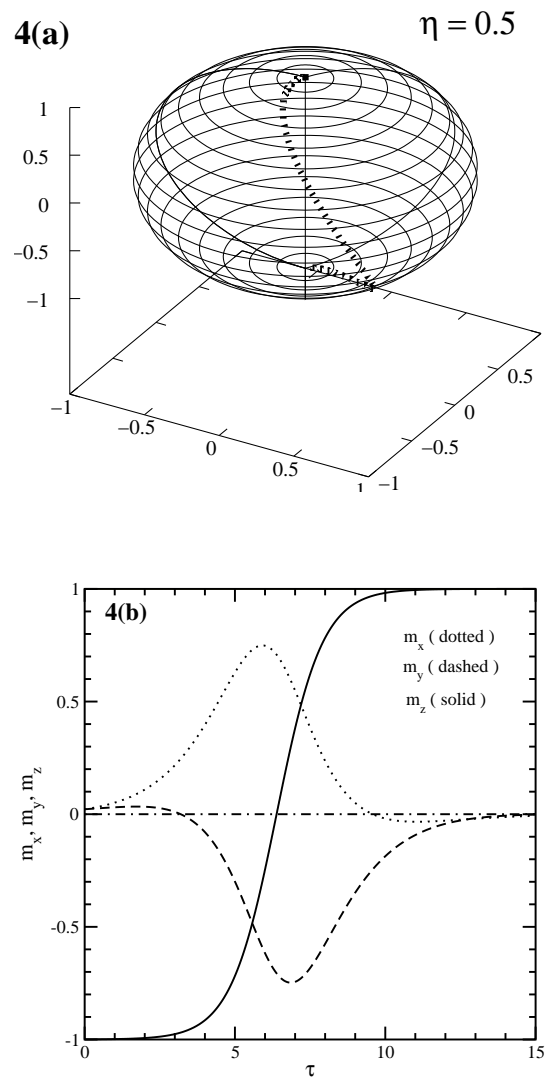}
\vskip 1.1cm
\includegraphics[height=2.6in]{fig4b.eps}
\caption{Plain reversal of the magnetic moment for $\eta=0.5$ and $a=1$.
Fig.~4(a) shows the locus (dotted curve)
of ${\bf m}$ and Fig.~4(b) is for $m_{i}(\tau)$ vs. $\tau$.
The initial direction of ${\bf m}$ is the same as in Fig.~3. Note that
there are no oscillations in $m_{x}$ and $m_{y}$. Thin circles define
a uni-sphere.
}
\end{center}
\end{figure}

\begin{figure}[tp]
\begin{center}
\includegraphics[height=2.6in]{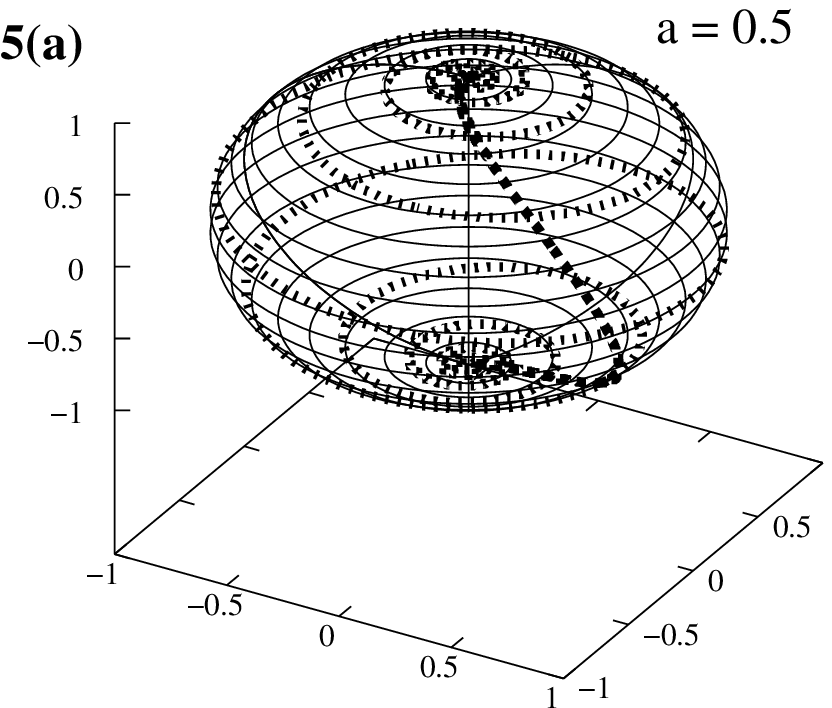}
\includegraphics[height=2.6in]{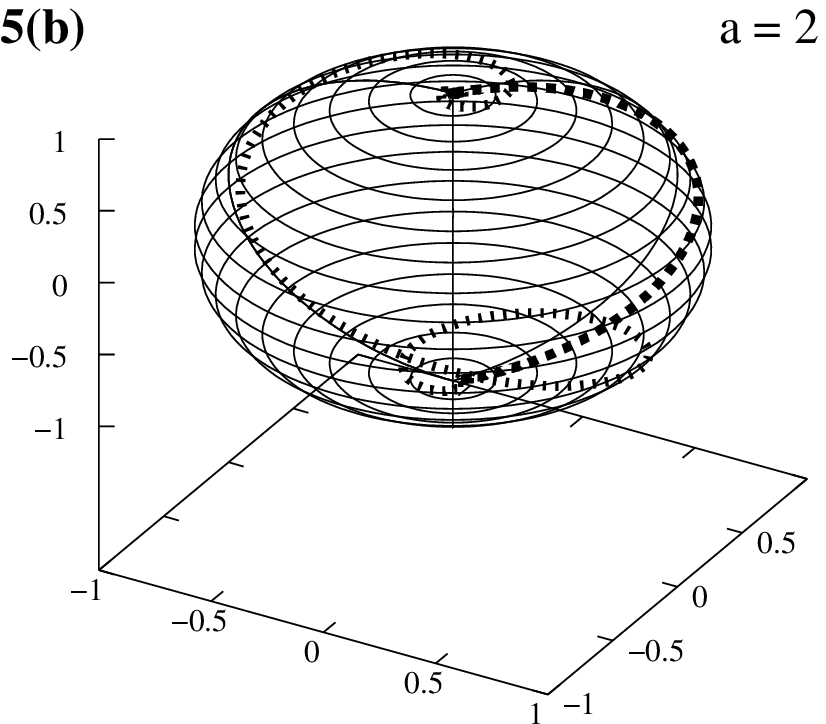}
\caption{Locus of ${\bf m}$ for $\eta=0.5$ (thick dotted curve)and $\eta=4$ (dotted curve). 
In Fig.~5(a), $a=0.5$
while $a=2$ in Fig.~5(b). Regardless of $a$, no precession occurs when $\eta=0.25$.
Thin circles define a uni-sphere.
}
\end{center}
\end{figure}

One can define the relaxation time $\tau_{0}$ of the reversal as an elapsed time
during the reversal between $\theta\simeq\pi$ and $\theta\simeq0$.
When $m_{z}\simeq1$, we can parametrize 
$\ln\left[1-m_{z}(\tau)\right]=c_{1}-c_{2}\tau/\tau_{0}$, where $c_{1}
\approx 8.9$
and $c_{2} \approx 13.6$. 
We found these values are independent of $\eta$ and $a$.
For given $\eta$ and $a$, we can determine $\tau_{0}$ by comparing
numerical results with $c_{1}-c_{2}\tau/\tau_{0}$. For example, $\tau_{0}\simeq
7.9$ for $\eta=0.9$ and $a=1$. 
In general, the smaller $a$ (or $l$) is, the longer $\tau_{0}$ is for a given $\eta$.
This can be understood because the wave function $|\psi_{tr}\rangle$ decays
faster if $l$ is shorter so that the spin transfer is relatively less effective
and, thus, it takes a longer time to reverse ${\bf m}$. 
In Fig.~6, we plot $m_{z}$ vs. $\tau$ for $\eta=4$ (main frame) and for $\eta=0.25$
(inset) with $a=0.5$ (solid) $1$ (dashed), and $2$ (dotted curve).
In this figure, we can see the 
relation between $a$ and $\tau_{0}$ mentioned above.
For $\eta\ge1$, as $a$ 
increases, weak precession occurs because $\tau_{0}$ decreases
as seen in the main frame of Fig.~6; in other words,
${\bf m}$ does not have enough time to precess strongly. 
We can also see such a behavior in Fig.~5 comparing $a=0.5$ and $a=2$
for $\eta=4$.

We plot $\tau_{0}$ vs. $\eta$ in Fig.~7 for
a given $a$. The relaxation time is evaluated
using the parameterization:
$\ln\left[1-m_{z}(\tau)\right]=c_{1}-c_{2}\tau/\tau_{0}$.
In the main frame, $a=1$ while in the inset $a=2$.
At $\tau=\tau_{0}$, $m_{z}(\tau_{0})\simeq0.99$ for all plots. 
Interestingly, $\tau_{0}$ is minimum at $\eta\simeq1$.
Therefore it is possible to estimate
the microscopic coupling parameter $J_{H}$ 
between an incoming spin and a magnetic moment
by measuring $\tau_{0}(\eta)$,
because $\tau_{0}$ has a minimum for a given mean free path.
\begin{figure}[tp]
\begin{center}
\includegraphics[height=2.6in]{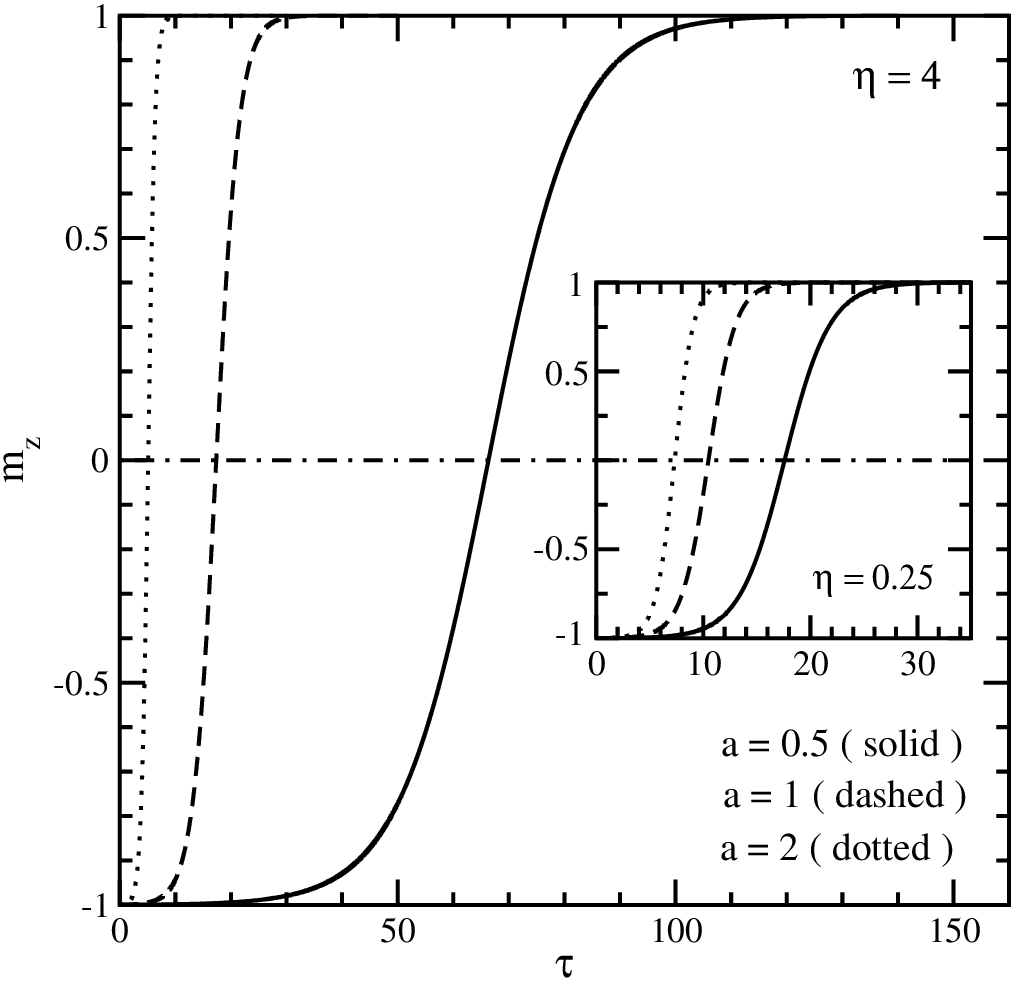}
\vskip 0.5cm
\caption{$m_{z}$ as a function of $\tau$. In the main frame, $\eta=4$ while
in the inset $\eta=0.25$ with $a=0.5$ (solid) $1$ (dashed), and $2$ (dotted curve).
}
\end{center}
\vskip 1cm
\begin{center}
\includegraphics[height=2.6in]{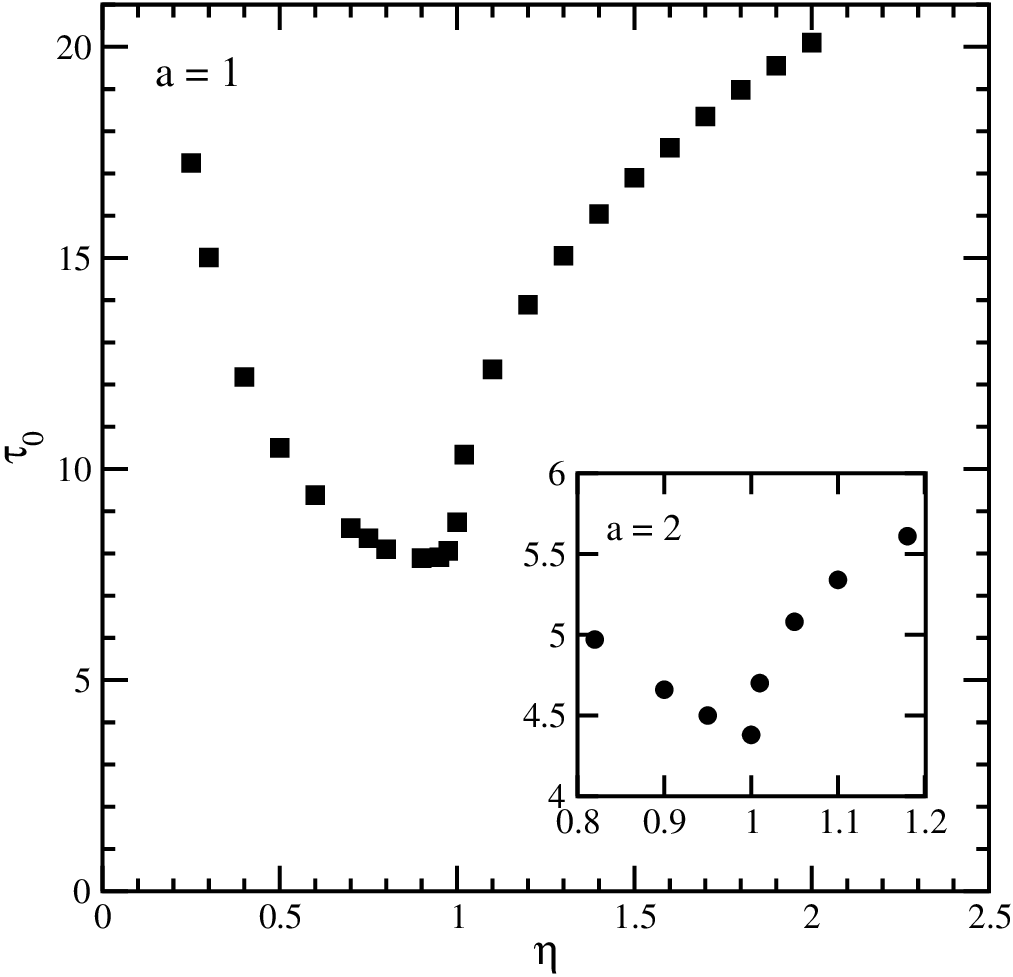}
\caption{The relaxation time $\tau_{0}$ vs. $\eta$.
In the main frame, $a=1$ while in the inset $a=2$.
$\tau_{0}$ has a minimum value at $\eta\simeq1$.
}
\end{center}
\end{figure}

\section{discussion and summary}

In this section we would like to discuss the adiabatic
approximation, which we tacitly used to study the dynamics of a magnetic moment.
First we summarize the procedure we followed.
We calculated ${\langle {\bf s}\rangle}$ using $|\psi (x)\rangle$;
namely, $\langle {\bf s}\rangle=(1/2)
\langle\psi (x)|{\bf\sigma}|\psi (x)\rangle$ 
for $x>0$ to solve
$
{d{\bf M}}/{dt}=2\gamma_{0}J_{H}
\left({\bf M}\times{\langle {\bf s}\rangle}\right)
$.
Here we mention that $|\psi (x>0)\rangle$ is obtained by considering
the Hamiltonian at a given time $t$ following Ref.\cite{berger}
Since the incoming wave function
$|\psi_{in}\rangle\sim|+\rangle$
is not an eigenstate of the Hamiltonian for $x>0$,  
we have a linear combination of $|+\rangle$ and $|-\rangle$
for $|\psi (x>0)\rangle$ and $|\psi (x<0)\rangle$. 
The matching conditions of wave functions at $x=0$
allow us to express the coefficients of the combination for $|\psi (x>0)\rangle$
in terms of ${\bf M}(t)$ (see Eqs.~(\ref{psi_tr}) and (\ref{coeffs})).
Now ${\langle {\bf s}\rangle}$ is
a function of ${\bf M}(t)$, and the time dependence of ${\langle {\bf s}\rangle}$
is given exclusively by ${\bf M}(t)$. This means that the time evolution
of the wave function for $x>0$ is not fully taken into account. 
In addition to the equation for ${d{\bf M}}/{dt}$, 
one can derive
the time derivative of the spin operator 
using $d{\bf s}/dt=i\left[H, {\bf s}\right]$:
\be
\frac{d{\bf s}}{dt}+\nabla\cdot{\cal J}=2J_{H}\left({\bf s}\times{\bf M}\right)\;,
\label{eq_spin}
\ee
where ${\cal J}$ is the spin-current tensor. It is obvious that
when we calculate an expectation value of ${\bf s}$ in Eq.(\ref{eq_spin})
we need to use $|\psi(x,t)\rangle$; 
$\langle {\bf s}\rangle_{t}=(1/2)\langle\psi(x,t)|\sigma|\psi(x,t)\rangle$,
where $|\psi(x,t)\rangle$ is obtained from $i\frac{d}{dt}|\psi(x,t)\rangle
=H|\psi(x,t)\rangle$. Rigorously speaking, one has to solve
the two coupled equations for ${\bf M}$ and ${\bf s}$ using
$|\psi(x,t)\rangle$ to calculate the expectation value of
${\bf s}$ and $\nabla\cdot{\cal J}$. 
However, if we compare Eq.~(\ref{mdot}) or (\ref{mdot2}) with Eq.~(\ref{eq_spin}),
we see that Eq.~(\ref{mdot2}) has a factor $1/S_{local}$ 
while Eq.~(\ref{eq_spin}) 
does not. This means that if we treat the magnetic moment semiclassically, i.e.
$S_{local}\gg1$, then the time scale of Eq.~(\ref{mdot2})
is much longer than that of Eq.~(\ref{eq_spin}). Therefore, the adiabatic
approximation is applicable to our analysis.

In summary,
we have studied the
effect of an incoming spin-polarized current on a local magnetic moment
in a magnetic thin film. We found that the spin torque occurs only when
the incoming spin changes as a function of time inside of the magnetic
film. If the incoming spin
remains parallel to its initial direction, no spin torque takes place.
This implies that some modifications
are necessary in a phenomenological model where the coefficient of the
spin torque term is a constant. Moreover, the coefficient is determined
by dynamics instead of geometrical details.
The magnetization reversal can be precessional as well as non-precessional
depending on the incoming energy of electrons in the spin-polarized current.
If the incoming energy is greater than the interaction energy $(J_{H}M_{0})$,
the magnetization precesses while reversing its direction. For
the incoming energy smaller than $J_{H}M_{0}$, the magnetization reversal is
non-precessional. We also found that the relaxation time associated with the
reversal depends on the incoming energy for a given mean free path. 
Our numerical
calculations imply the 
coupling between an incoming spin and a magnetic moment
$J_{H}$ can be estimated by measuring the relaxation time.

\begin{acknowledgments}
We thank Mark Freeman for interest and helpful discussions.
This work was supported in part by the Natural Sciences and Engineering
Research Council of Canada (NSERC), by ICORE (Alberta), and by the
Canadian Institute for Advanced Research (CIAR).
\end{acknowledgments}

\bibliographystyle{prb}

\end{document}